\documentclass{aa}
\usepackage{psfig}

\def\etal{{\it et al.}}
\def\ros{{\it ROSAT}}
\def\asec{\ifmmode ^{\prime\prime}\else$^{\prime\prime}$\fi}
\def\amin{\ifmmode ^{\prime}\else$^{\prime}$\fi}
\def\degs{\ifmmode ^{\circ}\else$^{\circ}$\fi}
\def\fdg{\hbox{$.\!\!^\circ$}}          
\def\fss{\hbox{$.\!\!^{\rm s}$}}        
\def\farcs{\hbox{$.\!\!^{\prime\prime}$}}  
\def\fd{\hbox{$.\!\!^{\rm d}$}}            

\def\rxj18{RX\,J1802.1+1804}

\begin{document}

   \thesaurus{06        
              (02.01.2, 
               08.02.3, 
               08.09.2, 
               08.14.2, 
               13.25.5  
                         )}

\title{The X-ray stream-eclipsing polar RX\,J1802.1+1804}

\author{J. Greiner \inst{1}\thanks{Present address: Astrophysical Institute
    Potsdam, An der Sternwarte 16, D-14482 Potsdam, Germany; jgreiner@aip.de}
    \and R.A. Remillard \inst{2}
    \and C. Motch \inst{3} }

\offprints{J. Greiner}

   \institute{Max-Planck-Institut f\"ur extraterrestrische Physik,
               D-85740 Garching, Germany 
          \and Center for Space Research, MIT, 77 Mass Ave, Cambridge, 
                MA 02139 U.S.A. \\
                email: rr@space.mit.edu
          \and Observatoire de Strasbourg, 11 rue de
                l'Universit\'e, F-67000 Strasbourg, France \\
                email: motch@astro.u-strasbg.fr          }  

   \date{Received  22 December 1997; accepted 31 March 1998}

   \maketitle

   \begin{abstract} We present X-ray and optical data of the new\-ly
discovered AM Her variable \rxj18.  This X-ray source was observed in the
\ros\ All-Sky-Survey in September 1990 and subsequently discovered as a
highly variable and soft X-ray source.  Follow-up pointed observations
confirmed the strong variability and revealed a periodic flux
modulation including a sharp and nearly complete eclipse of the X-ray
emission.  Based on the shape and duration of this eclipse as well as
the lack of optical eclipses we favour an interpretation in terms of a
self-eclipse by the accretion stream. The X-ray spectrum averaged over
the full period is dominated by soft emission below 0.5 keV. 
The ratio of  soft (blackbody) to hard
(bremsstrahlung) bolometric energy flux is 1670, distinguishing this object as
another example of a polar with an extreme strong excess in soft X-rays.


\end{abstract}

\section{Introduction}

AM Her type variables are accreting binaries in which the strong
magnetic field of the white dwarf controls the geometry of the
material flow coming from the late-type companion star. The radial
inflow of matter produces a shock front above the white dwarf surface
which gives rise to hard X-rays and polarized cyclotron radiation in
the IR to UV range.
In the simplest physical model (Lamb 1985) half of this shock
radiation intercepts the white dwarf surface and is reradiated as
thermal emission from the heated accretion spot in the UV and soft X-ray band.

X-ray observations, most notably with the \ros\ satellite, had and have
a great impact on the study of magnetic cataclysmic variables.
The rather high intensity, strongly variable and soft X-ray emission
allows to set up selection criteria (e.g. using the \ros\ data base) with
a high detection/identification rate which reflects in the fact that
about 80\% of all magnetic cataclysmic variables have been discovered
at X-ray wavelengths (Beuermann 1998). 

During the search for supersoft X-ray sources in the \ros\ (Tr\"umper
1983) All-Sky-Survey (RASS), we have found a bright X-ray source
showing strong scan-to-scan variability in the X-ray count rate as measured
in the PSPC.  Subsequent spectroscopic
observations re\-vealed a 15th magnitude cataclysmic variable located
near the X-ray position.  Additional pointed \ros\ PSPC observations
strengthened the evidence of a strongly modulated X-ray light curve as
well as of a very soft X-ray spectrum. These combined characteristics,
and particularly the supersoft X-ray spectrum, are typical signatures of
the AM Her subclass (Beuermann \& Schwope 1994).  Here, we report on the
results of our extensive \ros\ and optical observations performed to
study the accretion geometry in \rxj18\ in more detail. The identification 
as an AM Her system (polar) and some first details have been reported in 
June 1994 at the Abano-Padova conference on Cataclysmic Variables 
(Greiner \etal\ 1995a) after which the object got the variable star name
V884 Her. 
The results of quick follow-up observations of \rxj18\, (= WGA J1802.1+1804)
have also been communicated in Singh \etal\, 1995 (note the flux error; see
Greiner \etal\ 1995b) and Szkody \etal\ (1995). 
Recently, Shrader \etal\, (1997) reported the results of IUE observations.
Due to its soft
X-ray spectrum \rxj18\ has also been detected with the EUVE satellite 
(Lampton \etal\ 1997; = EUVE J1802+18.0).

\section{Observational data}

\subsection{X-ray observations}

\rxj18~ ( = 1RXS J180206.7+180438) was observed
over a 2.5 day period in September 1990 for a total of 716 sec
during the \ros\ All-Sky-Survey. These observations consist of 31 scans
with 9--29 sec exposure each, spaced in time by 96 min. (the ROSAT
orbital period).

A 2 ksec pointed \ros\ observation was performed on 1992 October 11 in
order to verify the strong variability and the dominant soft X-ray
excess. The results obtained from this observation motivated us to
propose a longer, 15 ksec pointed \ros\ PSPC observation which then was
performed in September 1993.  In order to minimize the varying
obscuration effects of the PSPC window support structure due to the
wobbling, \rxj18\ was by purpose observed at 39\amin\ off-axis angle during the
latter observation.  At this position in the detector, the FWHM of the
point spread function is 4\amin\ wide and thus the source photons are
spread over an area much larger than the 3\amin\ spacing of the
support structure wires.  Due to a commanding error this off-set option
was not implemented during the first 10\% of the 1993 exposure. Details
of all the \ros\ observations are given in Tab. \ref{obslog}.

\begin{table}
  \caption{Observation log of \rxj18}
  \begin{tabular}{crrc}
  \hline
  \noalign{\smallskip}
  \multicolumn{4}{c}{X-ray observations} \\
  \noalign{\smallskip}
  \hline
  \noalign{\smallskip}
  \hline
  \noalign{\smallskip}
     Date &  exposure~  & off-axis~~~ & cycles \\
  \noalign{\smallskip}
  \hline
  \noalign{\smallskip}
     Sep. 1990         &      716 sec~  &  0--55\amin~~~ &
                                           $\!\!\!\!$--13966...--13949$\!\!$ \\
     Oct. 11, 1992     &     2005 sec~  &  0\amin~~~ & --4269 \\
     Mar. 22, 1993     &      290 sec~  &  0\amin~~~ & --2060 \\
     Sep. 7, 1993      &   1\,239 sec~  &  0\amin~~~ & --50 \\
     Sep. 11/12, 1993  &  13\,477 sec~  &  39\amin~~~ & --2...12 \\
  \noalign{\smallskip}
  \hline
  \noalign{\smallskip}
  \hline
  \noalign{\smallskip}
  \multicolumn{4}{c}{Optical observations} \\
  \noalign{\smallskip}
  \hline
  \noalign{\smallskip}
  \hline
  \noalign{\smallskip}
     Date  & \multicolumn{2}{c}{Data type}    & Observatory  \\
  \noalign{\smallskip}
  \hline
  \noalign{\smallskip}
     Feb. 1, 1992      & \multicolumn{2}{c}{photometry}  &  OHP   \\
     Feb. 3, 1992      & \multicolumn{2}{c}{spectroscopy}  &  OHP  \\
     Jun. 3, 1992     & \multicolumn{2}{c}{photometry} &  MDM  \\
     Oct. 12, 1992     & \multicolumn{2}{c}{photometry}  &  MDM   \\
     Oct. 10--12, 1992 & \multicolumn{2}{c}{spectroscopy}  &  MDM   \\
     May 1/2/6, 1993   & \multicolumn{2}{c}{spectroscopy}  &  MDM   \\
  \noalign{\smallskip}
  \hline
  \end{tabular}
  \label{obslog}
\end{table}

\subsubsection{Position}

Although the optical identification of \rxj18\, had been achieved with
the position derived from the RASS, we improved the X-ray coordinates
by using the two pointed PSPC observations of Oct 11, 1992 and Sep. 7,
1993. Ignoring all X-ray photons below channel 25 to avoid ghost
images we derive a best fit X-ray position of R.A. (2000.0) = 18$^{\rm
h}$02$^{\rm m}$06\fss2, Decl. (2000.0) = 18$^\circ$04\amin42\asec\,
($\pm$15\asec).

   \begin{figure}[ht]
      \vbox{\psfig{figure=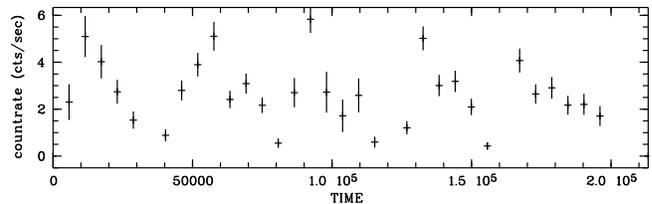,width=8.8cm,%
          bbllx=.8cm,bblly=5.3cm,bburx=19.2cm,bbury=11.05cm,clip=}}\par
      \caption[Survey]{Scan-to-scan X-ray light curve of RX J1802.1+1804.
        The duration of each measurement ranges between 9 and 29 sec.
        The obvious variability is the beat between the orbital period
        of the \ros\ satellite and the orbital modulation of the variable
        X-ray source.
      }
      \label{survlc}
   \end{figure}

   \begin{figure}
      \vbox{\psfig{figure=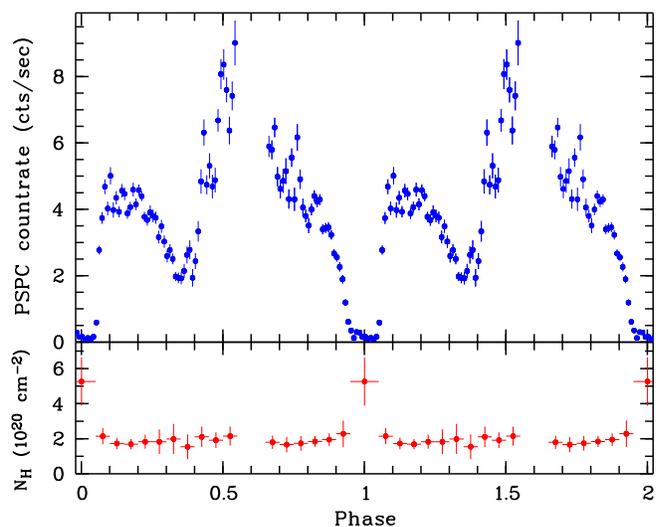,width=8.8cm,%
          bbllx=2.3cm,bblly=.9cm,bburx=17.2cm,bbury=13.1cm,clip=}}\par
         \vspace{-0.3cm}
      \caption[folded]{X-ray light curve of the total emission folded with the
          best fit X-ray period of 6780.65 sec = 113.0108 min. = 1.8835139 hr 
          = 0.07847977 days (with 
         1\,$\sigma$ statistical errors) and plotted twice in 1.1 min bins. 
         The gap around phase 0.55--0.65          are not exposed phase bins.
         The lower panel shows the variation of the absorbing column derived 
         by phase-resolved fitting of the X-ray spectra with fixed 
         $kT_{\rm bb}$ = 30 eV and $kT_{\rm th}$ = 20 keV (see paragr. 2.1.3).
      }
      \label{foldlc}
   \end{figure}

\subsubsection{Temporal variability}

Source photons have been extracted with a radius of 10 times the FWHM
of the \ros\ PSPC point spread function, excluding three faint nearby
sources (5\amin--8\amin\ away) in the off-axis pointing. The data from the
different pointings have been barycenter corrected and combined for
the subsequent analysis. The X-ray photons have been binned into 10 sec 
time slices, and during this binning process have been
background subtracted and corrected for vignetting using
standard EXSAS routines (Zimmermann \etal\ 1994).

 \begin{table}
 \caption{Heliocentric Julian dates of the mid-eclipses together with 
    the epoch (cycle counting) and deviation from phase zero 
    ($O-C$ = observed minus calculated values) implied by the ephemeris.}
  \begin{tabular}{ccr}
  \hline 
  \noalign{\smallskip}
  $T_{\rm ecl}$  & $O-C$ & cycle  \\
  (HJD 2400000+) & (10$^{-4}$ days) & \\ 
    \noalign{\smallskip} 
    \hline 
    \noalign{\smallskip}
         48146.26370$\pm$ 0.00405 & --19.2  & --13966 \\
         48146.73103$\pm$ 0.00405 & --33.4  & --13960 \\
         48147.13115$\pm$ 0.00405 &   22.5  & --13955 \\
         48147.59849$\pm$ 0.00405 &   8.4  & --13949 \\
         48907.28157$\pm$ 0.00023 &   3.3  &  --4269 \\
         49242.31274$\pm$ 0.00116 &   2.5  &   \,0 \\
         49242.39064$\pm$ 0.00035 &   0.2  &   \,1 \\
         49243.17597$\pm$ 0.00035 &   2.0  &    11 \\
         49243.25458$\pm$ 0.00035 &   6.8  &    12 \\
     \hline 
     \end{tabular}
     \label{omc}
     \end{table}

   \begin{figure}
      \vbox{\psfig{figure=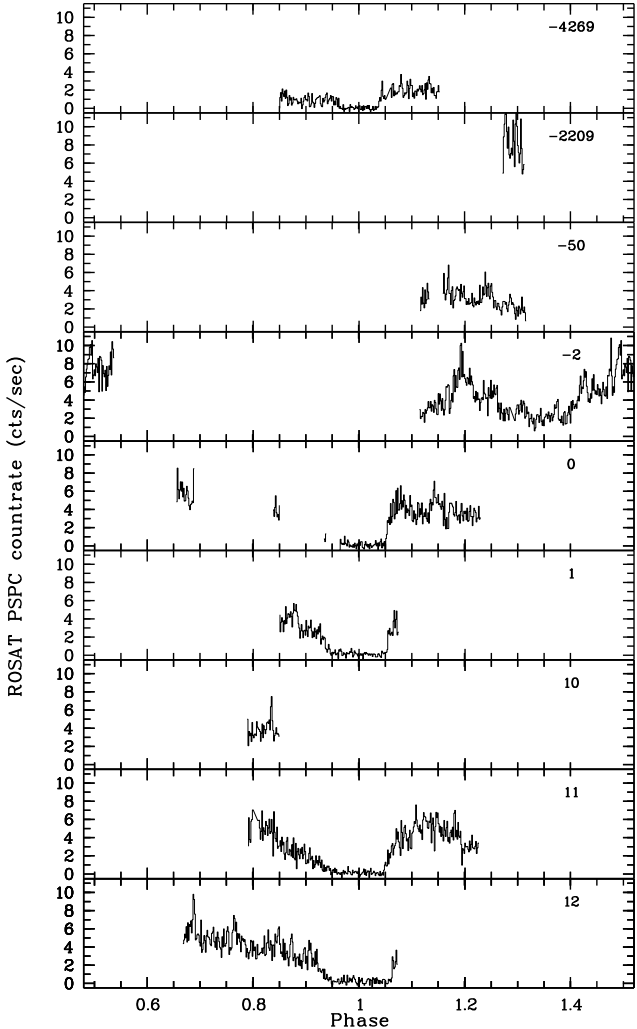,width=8.5cm,%
          bbllx=1.8cm,bblly=1.5cm,bburx=16.9cm,bbury=26.1cm,clip=}}\par
      \vspace*{-0.1cm}
      \caption[unfolded]{Individual pointed X-ray observations binned in 
          10\,sec intervals (over full energy range 0.1--2.4 keV) and 
          aligned in phase with the
          best fit X-ray period of 6780.65 sec. The number in each panel 
          gives the cycle number according to the ephemeris.
      }
      \label{unfoldlc}
   \end{figure}

Strong variability of the X-ray flux was already detected during the
All-Sky-Survey (see Fig. \ref{survlc}) and confirmed in the pointed
observations. The X-ray variations are periodic, and the light
curve is characterized by a double-wave per orbital cycle. In the
first minimum, which we consider as a periodic X-ray eclipse, the
source flux decreases to nearly zero. This is followed by a second
minimum occurring 40 min. later with a factor of 2 decrease in intensity.

In total we have covered 4 X-ray eclipses during the all-sky survey and
5 eclipses during the pointed observations. 
The fortunate circumstance of covering 4 eclipses spread over 17 cycles
during the survey and 4 eclipse egresses
spread over 12 orbit cycles during the Sep. 11/12, 1993 observation
allows us to phase all data together without any phase ambiguity.
The times of the mid-eclipses are listed in Tab. \ref{omc} together with
the O--C values of a linear regression analysis which yielded
the following eclipse (mid-eclipse times) ephemeris:

$$ T_{\rm ecl} {\rm (HJD)} = 2449242.3124(21) + 0.07847977(11) \times E $$

\noindent The numbers in parenthesis give the uncertainties in the last
digits and $T_o$ corresponds to mid-eclipse of the first eclipse in the
Sep. 11/12, 1993 observation (panel 5 in Fig. \ref{unfoldlc}).
The remaining error is mainly due to the varying eclipse length (see below)
and the not complete coverage of the eclipses (30 sec. scans) during the 
all-sky survey.

  \begin{figure}
    \vbox{\psfig{figure=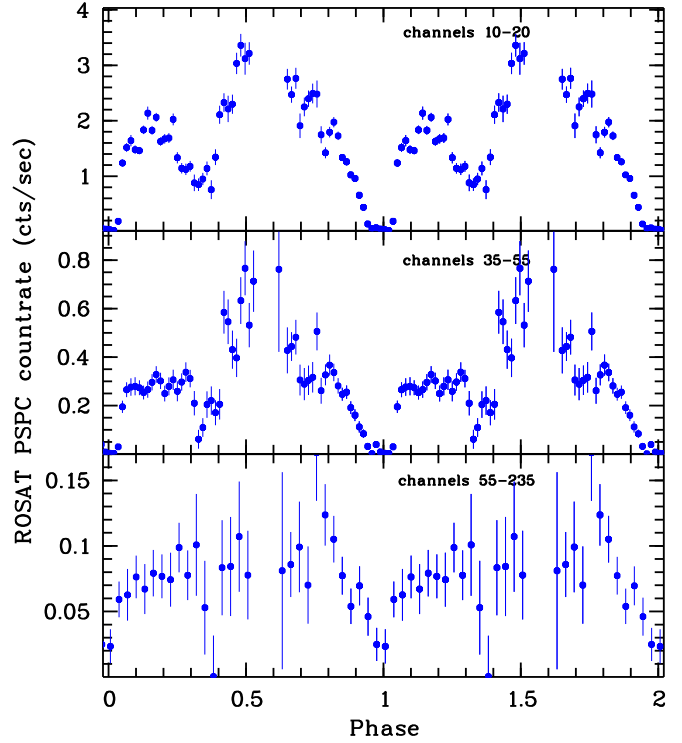,width=8.8cm,%
          bbllx=2.1cm,bblly=1.3cm,bburx=13.9cm,bbury=14.7cm,clip=}}\par
    \caption[folded3]{Folded X-ray light curve in different energy bins of the 
         PSPC as labeled in each panel (channel number divided by 100 
         gives roughly the energy in keV).
          }
      \label{foldlc3}
   \end{figure}

The FWHM of the eclipse is 17.0 min (or a phase interval of 0.15), and
the width of the second intensity drop is 14.7 min (0.13 in phase).
In addition, there are a number of pronounced intensity variations which
typically last a few minutes and reach amplitudes of 10\%--20\%.

   \begin{figure*}
      \vbox{\psfig{figure=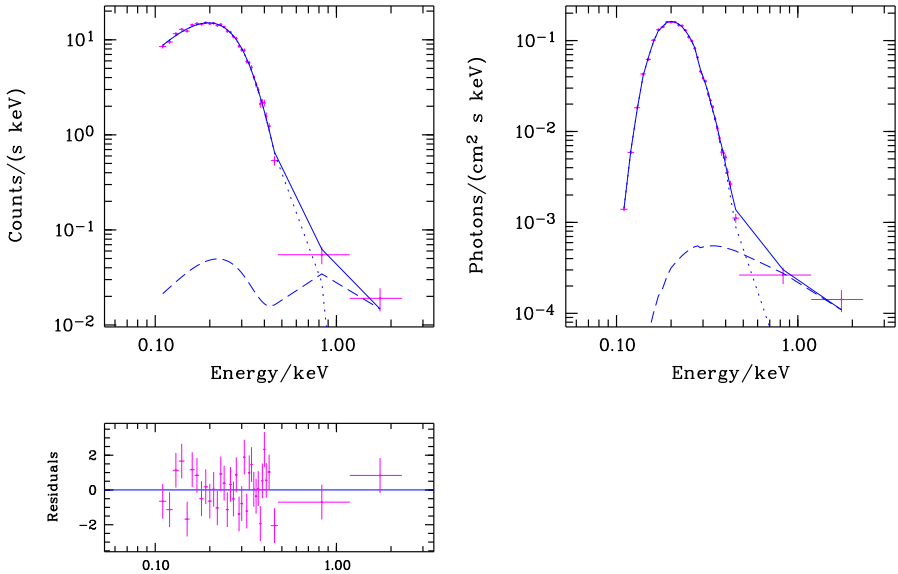,width=12.cm,%
          bbllx=2.4cm,bblly=6.7cm,bburx=19.4cm,bbury=17.6cm,clip=}}\par
      \caption[Spectrum]{X-ray spectrum of the emission averaged over the full
        period. 
        A blackbody plus thermal bremsstrahlung model has been fit to the data.
        With the temperature of the bremsstrahlung component fixed to 
        $kT_{\rm th} = 20$ keV the best-fit blackbody temperature is 
        $kT_{\rm bb} = 20\pm 15$ eV and the absorbing column 
        $N_{\rm H} =$ (3.2$\pm$1)$\times$10$^{20}$ cm$^{-2}$.
      }
      \label{spec}
   \end{figure*}

The light curve is strongly variable from orbit to orbit with the eclipse being
the only stable feature (Fig. \ref{unfoldlc}).  One particularly pronounced
difference is the strong flare around phase 0.2 during cycle --2
(Fig. \ref{unfoldlc}). The low intensity (especially before the eclipse)
during the Oct. 1992 observation (cycle --4269) and the high intensity during
the March 1993 observation (cycle --2209) are also noteworthy features.
The panels of Fig. \ref{unfoldlc} with the individually phased observations 
do not show the secondary minimum around phase 0.35 seen in 
the folded light curve (Fig. \ref{foldlc}), thus suggesting
that it is not a pronounced minimum but produced by the strong flare
in cycle --2.
We finally note that the eclipses are strongly variable in shape suggesting
an eclipse by the stream rather than that from a rigid body (see below).

\subsubsection{Spectral characteristics}

For the spectral analysis the X-ray photons have been background subtracted,
vignetting corrected and binned into amplitude channels using a constant
signal-to-noise ratio of 9$\sigma$. The X-ray spectrum of \rxj18\ 
(Fig. \ref{spec}) is dominated by soft emission below 0.5
keV.  In addition to this soft component there is faint but detectable
hard emission (above 0.5 keV) which could be explained by the usual
bremsstrahlung emission of AM Her variables. The temperature of this
bremsstrahlung component cannot be confined with our spectra (Fig. \ref{spec})
due to poor photon statistics at higher \ros\ energies and the restricted
spectral range of the PSPC, thus a direct estimate of the bremsstrahlung
component is possible only with ASCA data (Ishida \etal\ 1998).  
We therefore adopt a
temperature of $kT_{\rm th} = 20$ keV (and kept it fixed) for the spectral
fitting.  The best fit parameters for the mean phase-averaged spectrum
are a blackbody temperature of $kT_{\rm bb}$ = 20$\pm$15 eV and an absorbing
column of $N_{\rm H}$ = (3.2$\pm$1)$\times10^{20}$ cm$^{-2}$ (including
systematic errors). The count rate spectrum and the unfolded photon spectrum
are shown in Fig. \ref{spec} together with the residuals between model and
data. This absorbing column is about half of the total
galactic column in this direction (8.9$\times10^{20}$ cm$^{-2}$; Dickey \&
Lockman 1990) suggesting that at galactic latitude of b$^{\rm II}$=18\fdg8 
\rxj18\ may be at a distance of 100--200 pc.

The mean intensity of \rxj18\ in all \ros\ observations in the 0.1--2.4
keV band (excluding 0.1 phase units at the eclipse) is 3.4 cts/sec.  
The mean unabsorbed
bolometric blackbody and bremsstrahlung fluxes (again excluding the eclipse) 
are F$_{\rm bb}$=6.8$\times$10$^{-9}$ erg/cm$^2$/s and F$_{\rm
th}$=4.1$\times$10$^{-12}$ erg/cm$^2$/s, respectively (with some caution 
concerning the extrapolation of the soft component towards lower energies),
corresponding to luminosities of
L$_{\rm bb}$=7$\times$10$^{33}$ (D/100 pc)$^2$ erg/s and
L$_{\rm th}$=4.3$\times$10$^{29}$ (D/100 pc)$^2$ erg/s.
The ratio of photon flux of the blackbody to the bremsstrahlung component 
is 1670$\pm$150 (bolometric) and 86$\pm5$ in the 0.1--2.4 keV range, 
respectively (using the same assumptions as in Beuermann \& Burwitz 1995).
Even if we account for the fact that blackbody fits overestimate the
luminosity up to a factor of 2 as compared to white dwarf atmosphere
models (Williams \etal\ 1987, Heise \etal\ 1994), this soft excess would
still be large.

For the purpose of visualizing possible spectral variations with phase
we have plotted the folded light curve in three distinct energy ranges
(Fig. \ref{foldlc3}): (i) in channels 10--20, mostly affected by
absorption effects, (ii) in channels 30--55, the part of the blackbody
flux not affected by interstellar absorption and (iii) channels 55--235, the
high-energy part of the \ros\ spectrum which is dominated by the hard,
bremsstrahlung component. Two main characteristics are obvious: First,
there is no great difference between the two soft panels except the
relative depth of the intensity dips, suggesting that intrinsic absorption of
soft X-ray photons plays only a minor role in the X-ray modulations
outside of the dips.  Second, the main intensity peak in soft X-rays
is not accompanied by a corresponding peak above 0.55 keV. Moreover, the
hard photon intensity peaks at a different phase.

The eclipse light curve is clearly asymmetric, with the egress being
faster than the ingress by a factor 2--3 (see Fig. \ref{unfoldlc}). In
order to investigate possible absorption effects we have divided the
data into 19 phase bins and fitted the sum of a blackbody and thermal
bremsstrahlung model to the data, with the bremsstrahlung temperature
again fixed at 20 keV and the blackbody temperature fixed at the best
fit value of the phase-averaged spectrum (30 eV).  Thus, only the
absorption and the two normalizations were free fit parameters.  In
order to have more than 1000 photons in each phase bin, the bin size
covering the eclipse had to be chosen twice as large as the other
bins.  The result is shown in the lower panel of Fig. \ref{foldlc} and
shows more or less the same absorbing column for all phase intervals
except the eclipse interval. 
Thus, we draw the following two conclusions: 
(1) The asymmetry in the eclipse light curve can not be explained by 
increased cold absorption during the ingress as compared to the egress.  
(2) The mean non-zero flux observed during eclipse is
more  absorbed than the emission outside eclipse by a factor of
2--3.  This absorption alone could cause a drop of the X-ray count
rate in the PSPC by a factor of 6--10 during this phase interval!
However, cold absorption of that amount should only reduce the PSPC
count rate below about channel 25, whereas Fig. \ref{foldlc3} reveals
that the drop in intensity occurs at basically all energies. In order to
effectively absorb photons up to channel 100 would require a cold
absorbing column (of the stream) of 10$^{22}$ cm$^{-2}$ incompatible with the 
pulse height spectrum during this phase interval. Thus, we conclude that the
intensity drop around phase 0 (--0.05--0.05) is not exclusively due to
cold absorption but needs  either partial covering or warm absorption.

   \begin{figure}
      \vbox{\psfig{figure=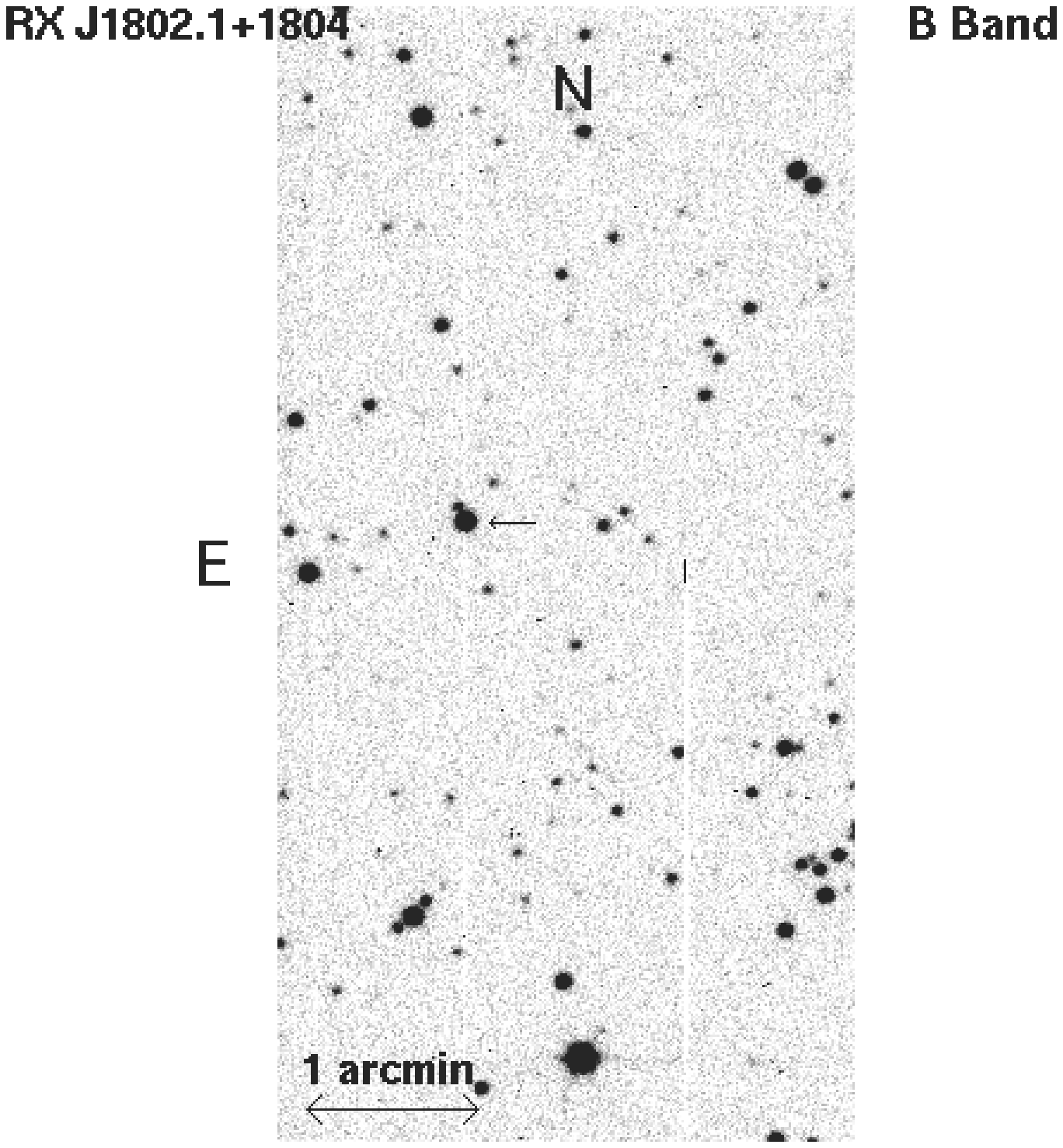,width=8.8cm,%
          bbllx=5.4cm,bblly=11.3cm,bburx=14.3cm,bbury=25.6cm,clip=}}\par
      \caption[CCD blue]{CCD blue image of \rxj18 with the optical
        counterpart marked by an arrow. North is up and east to the left. }
      \label{ccdima}
   \end{figure}

\subsection{Optical data}

   \begin{figure}
      \vbox{\psfig{figure=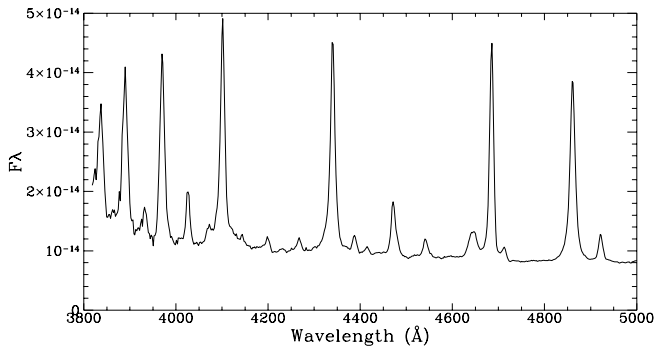,width=8.8cm,%
          bbllx=.6cm,bblly=4.5cm,bburx=20.2cm,bbury=15.cm,clip=}}\par
      \caption[opt. spectrum]{Mean optical spectrum of \rxj18\,
         (average of the May 1993 data)
         showing the typical signatures of AM Her variables.
          Flux units are erg/cm$^2$/s/\AA. }
      \label{ccdspec}
   \end{figure}

As noted above, \rxj18 is optically identified with a V=15 mag
cataclysmic variable. CCD imaging of the X-ray position was performed
with the 1.2 m telescope at Observatoire de Haute-Provence on February
1, 1992.  The images were flat-fielded using standard MIDAS routines.
A 3 min. B band image is shown in Fig. \ref{ccdima}, providing a
finding chart for the optical counterpart. The optical position is
measured as: R.A. (2000.0) = 18$^{\rm h}$02$^{\rm m}$06\fss4,
Decl. (2000.0) = 18$^\circ$04\amin43\asec ($\pm$1\asec).  

   \begin{figure}[thb]
      \vbox{\psfig{figure=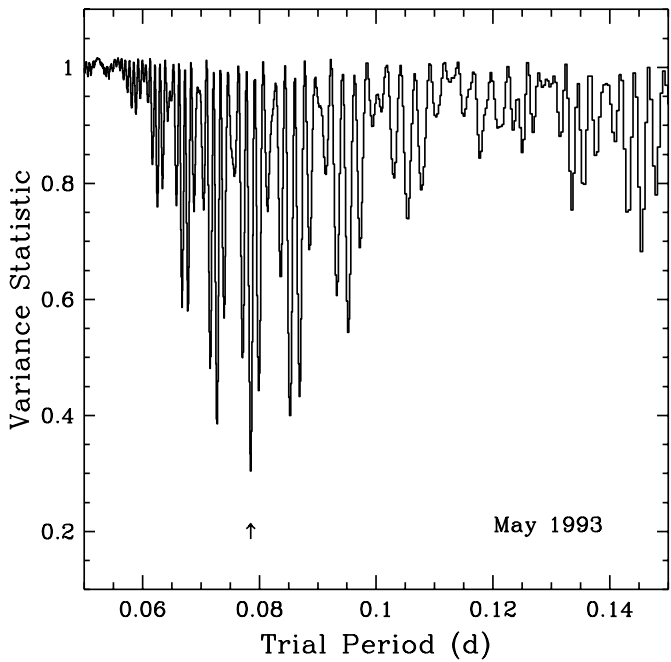,width=8.8cm,%
          bbllx=3.4cm,bblly=8.cm,bburx=17.4cm,bbury=22.cm,clip=}}\par
      \caption[periodogram]{Periodogram of the radial velocity curve
          derived from the May 1993 data. The best period at 0\fd078479
       (6780.6 sec) is marked by an arrow.}
      \label{ccdper}
   \end{figure}

Spectroscopic identification observation of the 15 mag. optical counterpart 
has been performed within the Galactic Plane Survey (Motch \etal\ 1991).
The optical spectrum of \rxj18\ (Fig. \ref{ccdspec}) 
shows high-excitation characteristics,
including broad emission lines of H, HeI and HeII, with strong
contributions of He II relative to He I and the Balmer series.

Phase-resolved spectroscopic observations were ob\-tai\-ned with
the 1.3 m McGraw Hill telescope at MDM Observatory at Kitt Peak,
Arizona on 1992 October 10, 11, and 12, and again on 1993 May 1, 2,
and 6.  The observations were made with the Mark III CCD spectrograph
using a 600 line/mm grism that yielded 2.8 \AA\ spectral resolution
(FWHM) over the wavelength range of 3800-5100 \AA.  The total exposure
times were 8 hr during 1992 October and 12.6 hr during 1993 May.  The
slit width was 3\farcs5, and all of the observations were autoguided.

   \begin{figure}
       \vbox{\psfig{figure=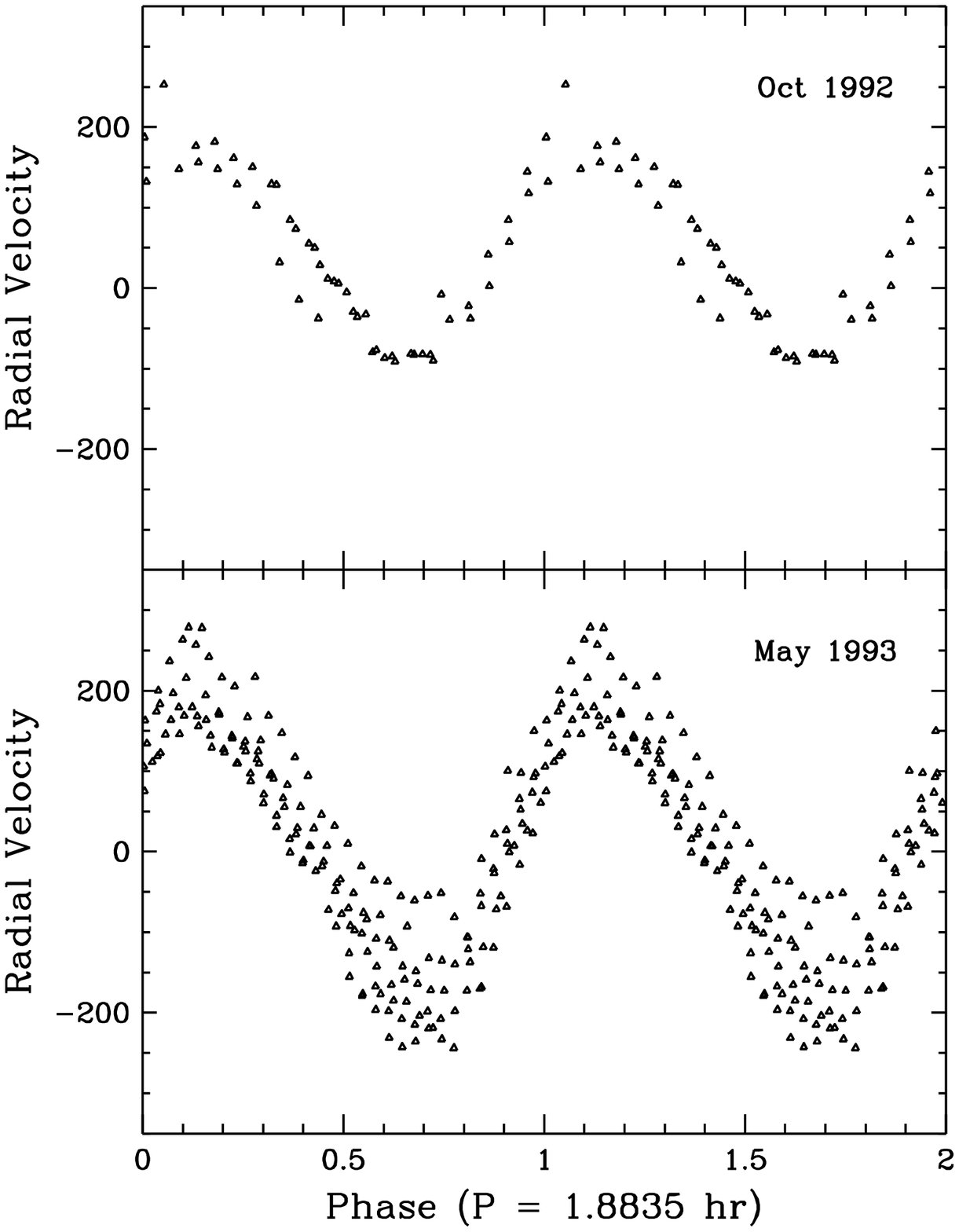,width=7.45cm,%
          bbllx=2.7cm,bblly=3.9cm,bburx=18.3cm,bbury=23.6cm,clip=}}\par
      \caption[rv]{Radial velocity curves of \rxj18\, during two epochs 
          using all strong emission lines. The velocity profile is stable
          over the 8 months period. }
      \label{rv}
      \vspace*{0.1cm}
      \vbox{\psfig{figure=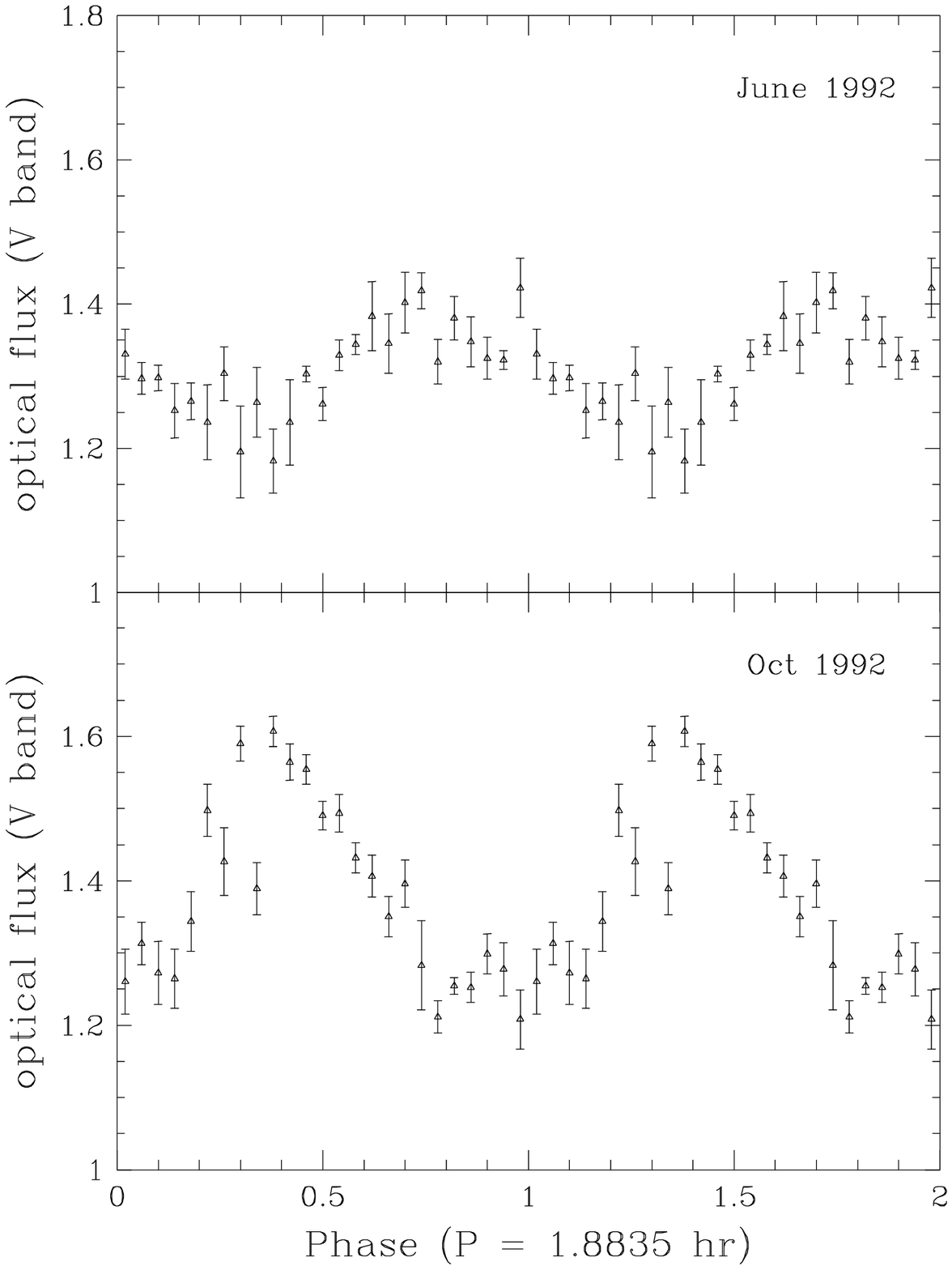,width=7.45cm,%
          bbllx=3.2cm,bblly=3.9cm,bburx=18.2cm,bbury=23.7cm,clip=}}\par
      \caption[rv]{Photometric observations of \rxj18\, of two epochs 
        folded with our best-fit ephemeris. Note that the flux scale is linear
        and is normalized to the mean intensity of the nearby comparison star.
        While the October 1992 show the typical profile expected for a
        stream-eclipsing polar, the June 1992 data show considerable 
        intensity variations in phase, shape and amplitude.  }
      \label{phot}
   \end{figure}

Data reductions were performed using the IRAF software package
distributed by NOAO.  The mean spectrum (see Fig. \ref{ccdspec}) was computed
for each observing run, and this result then served as the template for
measuring the radial velocities for individual spectra which were
obtained every 200 s.  The time series of radial velocities were then
analyzed for periodicities using a variance minimization technique
described by Stellingwerf (1978).  A significant period was found for
both observing runs.  The superior result was derived during 1993 May
(Fig. \ref{ccdper}), with a period of 0.078479 $\pm$ 0.00007 days 
(6780.6$\pm$6 sec).
This period coincides within the errors with the period derived from
the X-ray data.
Assuming this value as the binary period, we folded the velocities
(Fig. \ref{rv}) and fit the results to a sine wave, yielding velocity
semi-amplitudes, K = 81.0$\pm$7.2 km/s in 1992 October, and K =
154.5$\pm$6.0 km/s during 1993 May.  These velocity differences suggest
significant evolution in the accretion geometry between these two
epochs.  For example, there may have been changes in the proportion of
the accretion rate travelling to the two magnetic poles.

We have determined equivalent widths of some emission lines from these
May 1993 spectra, e.g. 61.3 \AA\ (H$\beta$), 51.3 \AA\ (H$\gamma$),
51.9 \AA\ (HeII 4686 \AA), and 13.2 \AA\ (HeI 4471 \AA). It is worth not only
to mention the large equivalent widths but also the high ratio of
He II/H and He II/He I.

The photometric observations revealed 0.3 mag variations, but there
were no optical eclipses. Folding the data with the above ephemeris gives
the curves as shown in Fig. \ref{phot}. During October 1992, the low 
intensity part occurs around the phase of the X-ray eclipse while the maximum
of the high intensity part is about 90\degs\ later. This suggests that
the emission is dominated by cyclotron emission and the modulation is due
to its beaming perpendicular to the magnetic field lines.
During June 1992 the optical light curve is distinctly different:
the phasing seems to be shifted, the amplitude is much smaller (about 0.1 mag)
and the shape is nearly symmetric. The maxima in the optical continuum also
vary in amplitude from cycle to cycle, but they consistently appear near or 
just after the time at which the radial velocity crosses the midpoint from 
red-shifted to blue-shifted phases.

A check of 48 astrograph plates (F series; exposure time typically 30--50 min.)
taken at Sonneberg Observatory
between 1930--1932 and 1949--1968 showed the object to be either constant near
the limiting magnitudes of the plates (amplitude smaller than 0.2 mag)
or invisible on plates of poorer quality (Wenzel 1994).

\section{Discussion}

\subsection{AM Her type classification}

The coincidence of periods derived from the X-ray and optical data
and the stability of the eclipse feature over more than 5 years 
suggest that the binary system is synchronized.
The evidence of an X-ray eclipse without a corresponding optical
eclipse together with the shape and variations with time of the
X-ray eclipse imply that (1) the optical and X-ray radiation originate from
different emission regions and (2) the X-ray emitting region is
confined to a small region near the white dwarf.  The localized X-ray
emitting region strongly indicates a magnetic CV subclass, since the X-ray
emission from an inner accretion disk would only show eclipses when
the disk is occulted by the companion star -- and such eclipses are
evident in both X-ray and optical light.

The supersoft X-ray spectrum further suggests an AM Her / `polar'
subtype (rather than `intermediate polar' type), since the association
between AM Her objects and luminous emission in soft X-rays is well
established.  We note, however, that the discovery of three
intermediate polars with strong soft X-ray emission has changed this
clear association (Haberl and Motch 1995).  However, the consistency
of singular optical and X-ray periods in the case of \rxj18 indicates
the likelihood of spin / orbital synchronous rotation, another classic
property of AM Her binaries. Finally, the large equivalent widths of the
optical emission lines also suggest a polar nature.

An independent confirmation of the polar classification comes from 
polarimetric observations (Szkody \etal\
1995). The circular polarization in the 5700--7700 \AA\ range reaches a 
maximum of 4\% in two peaks separated by a minimum which occurs at the time
of photometric minimum.

\subsection{Geometric configuration}

The relatively short time scales observed for X-ray eclipse ingress,
egress, and duration argue against the interpretation of eclipses as
the disappearance of the X-ray emitting region behind the limb of the
rotating white dwarf. We therefore interpret the eclipse as a
self-eclipse of the X-ray emitting region by the accretion
stream. X-ray spectral evidence for increased absorption column
associated with the eclipse and the variable eclipse length provide 
further support for this conclusion.

Although one would expect the optical light curve to be more complex
than the X-ray light curve due to several reasons (cyclotron radiation
is beam-modulated and can be self-eclipsing; the recombination
component is modulated by projection of the partially optically thick
stream; additional light may be seen from the illuminated side of the
secondary), the observed optical light curve is surprisingly smooth.
Despite the stream-eclipse of the accretion spot in X-rays, the optical
light is not affected due to its probable origin from a more extended region,
including a portion of the accretion column upstream from the X-ray
emitting area. The lack of optical eclipses suggests an inclination of
$i < 78$\degr\ (assuming a typical M5/6 companion).

During mid eclipse, there is still detectable emission not only in the
total flux, but also in the soft X-ray component. Thus, it seems
possible that the accretion spot is not fully eclipsed or we see
another, uneclipsed emission component.

The presence of X-ray emission at all phases implies that 
$i + \beta <$ 90\degr\ (with $\beta$ being the colatitude of the accreting
magnetic pole).
Combining this with the condition of stream eclipse ($i > \beta$) we
derive the following constraints on $i$ and $\beta$:

$$ 45\degr < i < 78\degr\ \hspace{0.5cm} {\rm and} \hspace{0.5cm} 
0\degr < \beta < 45\degr $$

The X-ray light curve is very complex, and there are large differences
seen when comparing observations well separated in time.  This might
suggest that the density and/or the size of the accretion stream vary
with time and thus cause changes in the size and/or vertical extent of
the accretion spot. In this regard, we note that the eclipse duration
is not constant.  Also, there are large changes in the characteristics of a 
secondary minimum, as noted in Section 2.1.2 and shown in Fig. \ref{foldlc3}.
Furthermore, the eclipse length seems to be correlated with the
intensity of the X-ray emission before and after the eclipse. During
the Oct. 1992 observation (cycle --4269 in Fig. \ref{unfoldlc}) with its
X-ray low-state, the eclipse length is only 0.07 phase units, whereas
it is 0.1 during the high-state in Sep. 1993 and even 0.12 in the last
eclipse observed.  A higher X-ray intensity may be caused by an
increase in the rate of mass transfer, perhaps leading to a larger
width of the accretion stream, which then increases the duration of
the X-ray eclipse.

A significant larger ingress (5--7 min.) than egress (2--2.5 min.) is
measured. Such asymmetry is possible if the impact area is not a circle
but an arc due to the coupling of the stream to nonpolar field lines
(see e.g. Beuermann \etal\ 1987 and their notion of ``X-ray auroral oval''. 
Alternatively, the accretion stream can be imagined to 
impact the white dwarf surface not from a perpendicular direction
thus forming an elongated footprint on the white dwarf surface. 
The slow fall of X-ray intensity into the eclipse minimum then can be
understood as being due to two effects: first, the projected area of
the footprint decreases due to the rotation of the white dwarf and secondly, 
the stream starts to occult the 
footprint area. This is consistent with the observed X-ray spectral property 
that the absorbing column does not change during the start of the slow
fall into eclipse minimum (phase interval 0.0--0.1, see Fig. \ref{foldlc}), 
but only during the eclipse (phase interval 0.1--0.2).
Such behaviour is also consistent with the interpretation of
the variation of the polarization being due to varying aspects of an
emission region extended in longitude (Szkody \etal\ 1995).
A size estimate of this elongated emitting region beyond an axis ratio of
about 2--3:1 is hard to evaluate since
the relative sizes of the emitting region and the stream are important. 

The X-ray eclipse occurs about 35\degr\ after the blue-to-red crossing of 
the radial velocity curve. In our interpretation of a stream-eclipsing 
geometry this is consistent with the picture that the emission lines 
are produced in the ballistic part of the accretion stream.
We note, however, that the radial velocity amplitude is lower than one 
would expect as the maximum free fall velocity.

The best fit blackbody temperature and the normalization transform into a
corresponding blackbody radius of the emission region of 750 (D/100 pc) km.
Bearing in mind the above mentioned elongation this is understood as the
radius of a circle which has the same area as the elongated emitting region.
If true, this implies a stream diameter near the white dwarf surface
of the order of 850--1050 km.

Besides the herewith presented \rxj18\ several other polars have been found 
where the accretion stream was identified as the cause of narrow absorption 
dips in the EUV/X-ray light curves: 
EF Eri (Patterson \etal\ 1981, Beuermann \etal\ 1991),
UZ For (Warren, Vallerga and Sirk 1995),
EK UMa (Clayton \& Osborne 1994),
QS Tel (Beuermann \& Thomas 1993, Buckley \etal\ 1993, Schwope \etal\ 1995),
HU Aqr (Schwope \etal\ 1993, Hakala \etal\ 1993, Schwope \etal\ 1997),
QQ Vul (Beardmore \etal\ 1995) and 
V2301 Oph (= 1H1752+081) (Hessman \etal\ 1997).
Another stream-eclipsing polar may be AX\,J2315--592 which has recently been 
found from ASCA observations (Misaki \etal\ 1996, Thomas and Reinsch 1996).
The EUVE light curves of UZ For show that both the phase and
the amplitude of the dips vary over the 3-day observation suggesting 
variations in the density and position of the accretion stream similar to what
we observed in the \ros\ light curves of \rxj18.
Also, the X-ray light curve of QQ Vul as observed with ROSAT is surprisingly
similar to that of \rxj18\ (as given in Fig. \ref{foldlc}).

\subsection{Soft X-ray excess}

As in several other AM Her binaries, the soft X-ray luminosity of
\rxj18 is larger than the hard one,
a fact known as `soft X-ray problem' (Rothschild \etal\ 1981).
Kujpers \& Pringle (1982) proposed that
non-stationary accretion of dense blobs can heat the photosphere from below.
This implies that high magnetic field systems should have a weak bremsstrahlung
component, which is proved by a comparison of soft-to-hard
flux ratios with measured magnetic field strengths
in different AM Her systems (Beuermann \& Schwope 1994).
According to this correlation a magnetic field larger than 40 MG 
is suggested for \rxj18\ which has one of the highest ratios of soft/hard
emission among polars. This would produce a very impressive Zeeman
split in the H absorption lines coming from the white dwarf as well
as strong cyclotron humps (depending on the viewing angle). However,
since we observed the system ``only'' in its robust state of accretion, the
optical spectrum is strongly dominated by the accretion stream
making the determination of the magnetic field impossible.

\section{Summary}

We have identified the \ros\ all-sky-survey X-ray source \rxj18\ as a V=15 mag.
AM Her type object with a period of 6780.65$\pm$0.01 sec
(1.8835145$\pm$0.000003 hr). While the
optical light shows a smooth sinusoidal intensity variation, the X-ray
light curve exhibits a double wave structure with a nearly complete
eclipse and a broad dip about 0.35 phase units later.
We interpret the eclipse as being due to the self-occultation of the emission 
region by the accretion stream.

\rxj18\ is one of the X-ray brightest stream-eclipsing AM Her variables and
the study of this source allows a deep insight in the geometry of this
binary system. Further phase-resolved spectroscopy and polarimetry 
during a possibly upcoming optical low state should eventually reveal the 
magnetic field geometry.

\begin{acknowledgements}
JG is supported by the Deutsche Agentur f\"ur Raumfahrtangelegenheiten
(DARA) GmbH under contract numbers FKZ 50 OR 9201 and 50 QQ 9602\,3.
JG acknowledges substantial travel support from DFG GR 1350/6-1 for a
visit of MIT where this work was completed. RR acknowledges partial
support from NASA grant NAG5--1784.  The \ros\ project is supported by
the German Bundes\-mini\-ste\-rium f\"ur Bildung, Wissenschaft,
Forschung und Technologie (BMBF/DARA) and the Max-Planck-Society.
This research has made use of the Simbad database, operated at CDS,
Strasbourg, France.
\end{acknowledgements}

\end{document}